\documentclass[preprint,prl,aps,floats]{revtex4}
\usepackage{graphicx}
\usepackage{amsmath}
\begin{document}
\title{Structure Calculations Without Effective Interactions}
 \author{J.~A. Secrest$^1$, J.~M. Conroy$^2$ and  H.~G. Miller$^{2}$ }

 \affiliation{$^1$Department of Physics and Astronomy, Georgia Southern University, Armstrong Campus, Savannah, Georgia, USA}

 \affiliation{$^2$Department of Physics, State University of New York at Fredonia, Fredonia, New York,
USA}

\date{\today}
\begin{abstract}
Good approximate eigenstates of a Hamiltionian operator which poesses a point as well as a continuous spectrum have beeen obtained using the Lanczos algorithm. Iterating with the bare Hamiltonian operator yields  spurious solutions which can easily be identified. The rms radius of the ground state eigenvector, for example, is calculated using the bare operator.
\end{abstract}
\maketitle

The Hamiltonian operators in nuclear and atomic physics are usually only 
bounded from below and possess a point as well as continuous spectrum. Unlike 
in atomic physics where in a properly chosen basis, the Hamiltonian matrix is 
diagonally dominant this is definitely not the case in nuclear physics.
The generally accepted procedure for determining the lower-lying eigenstates in 
this case  is to first map the Hamiltonian operator onto a 
finite dimensional subspace space spanned by a number of $\mathcal{L}^2$
basis states, {\it{e.g.}} to  construct an effective Hamiltonian 
operator\cite{BH58,SW73,EO77,CCGIK09} which must then be diagonalized.
This procedure is fraught with difficulties as initially  the effective 
Hamiltonian operator is non-linear, non-hermetian and  upon diagionalization 
guarantees at best
only good approximate eigenvalues of the lowest lying eigenstates.  The 
eigenstates of the effective Hamiltonian are the projection of the exact 
eigenstates onto the subspace.  Hence to calculate any other observable the 
corresponding operator must also be mapped on the same subspace.

Consider an   operator, $\hat{H}$, 
which possesses  a continuum (or continua) as well as a point spectrum. The
subspace spanned  by 
its bound state eigenfunctions, $\mathcal{H_B}$,  is by itself certainly not
complete.
As the composition of this space is generally not known beforehand, a set of
basis states which is complete and spans a space, $\mathcal{F}$, is chosen to
construct a matrix representation of the operator, $\hat{H}$, to be
diagonalized.  Mathematically this corresponds to projecting the operator 
$\hat{H}$
onto the space $\mathcal{F}$. Clearly the eigenpairs obtained from
diagonalizing 
the projected operator, $\hat{H}_P$,  need not all be the same as those of the
operator $\hat{H}$.
However, because the set of basis states is complete, any state contained in
$\mathcal{H_B}$ can be expanded in terms of this set of basis states. Hence 
$\mathcal{H_B}$ may also be regarded as a subspace of $\mathcal{F}$ and the
complete
diagonalization of  $\hat{H}_P$ will yield not only the exact eigenstates of
$\hat{H}$
but additional spurious eigensolutions.  Note these spurious eigenfunctions are
eigenfunctions of $\hat{H}_P$   but not  of $\hat{H}$. Furthermore in this case
the Rayleigh-Ritz bounds discussed in the paper by Krauthauser and
Hill\cite{KH02}
apply now to the eigenstates of $\hat{H}_P$.

It is interesting to note that the same problem occurs in the Lanczos
algorithm\cite{L50,D98} when it is applied to operators which possess a bound 
state
spectrum  as well as a continuum\cite{AM03}.  This is not surprising as the
Lanczos algorithm can also be considered as an application of the Rayleigh-Ritz
method\cite{P80}. In this case an orthonormalized set of Krylov basis vectors 
is used to construct iteratively a matrix representatiion of the operator which 
is 
then diagonalized.  Again  spurious states can occur for precisely the same
reasons given above.
In this case we have proposed identifying the exact bound states in the
following manner\cite{AM03}. After  
each iteration,  for each of the converging eigenpairs ($e_{l \lambda}$,$|e_{l 
\lambda}\rangle$), $\Delta_{l \lambda}=|e_{l \lambda}^2-<e_{l 
\lambda}|\hat{H}^2|e_{l \lambda}>|$ (where $l$ is the iteration number) is 
calculated and a determination is made  as to whether  $\Delta$ is converging  
toward zero or not.  For the exact bound states of $\hat{H}$,  $\Delta$ must be 
identically zero while the  other eigenstates states of the projected operator 
should converge  to some non-zero positive   value. This method has been
successfully implemented to identify spurious states in
non-relativistic\cite{AM03}
as well as relativistic\cite{AMY09} eigenvalue problems. A similar procedure can
be implemented in any Rayleigh-Ritz application.  One simply must check to see
whether the eigensolutions from the diagonalization  of $\hat{H}_P$  are also
eigensolutions of $\hat{H}^2$.

To illustrate this a simple example is presented. Consider the following 
Hamiltonian operator

\begin{equation}
 \hat{H}= \hat{H_0} + \hat{V} 
\end{equation}
where $H_0$ is the harmonic oscillator Hamiltonian  with $\hbar\omega=1$  and 
\begin{equation}
 V= V_0\exp{-a r^2}
\end{equation}
with $V_0=-11$ and $a= 1 $.
 Rather than simply attempt to 
obtain a coordinate state representation of the eignstate\cite{AM03},  approximate 
eigenstates have been back expanded in a basis as is the case in more realistic many body calculations
calculations.
For a single particle the Hamiltonn matrix in an s wave 
oscillator
basis
\begin{equation}
 \phi(x)= \frac{1}{\sqrt{2^v\sqrt{\pi} v!}}H_v(x)e^{-x^2/2}
\end{equation}
where $H_v(x)$ are the Hermite polynomials.
an $n$-dimensional representation of $\hat{H}$ can easily be constructed. 

\begin{figure}[h]
  \begin{center}
     \includegraphics[height=60mm]{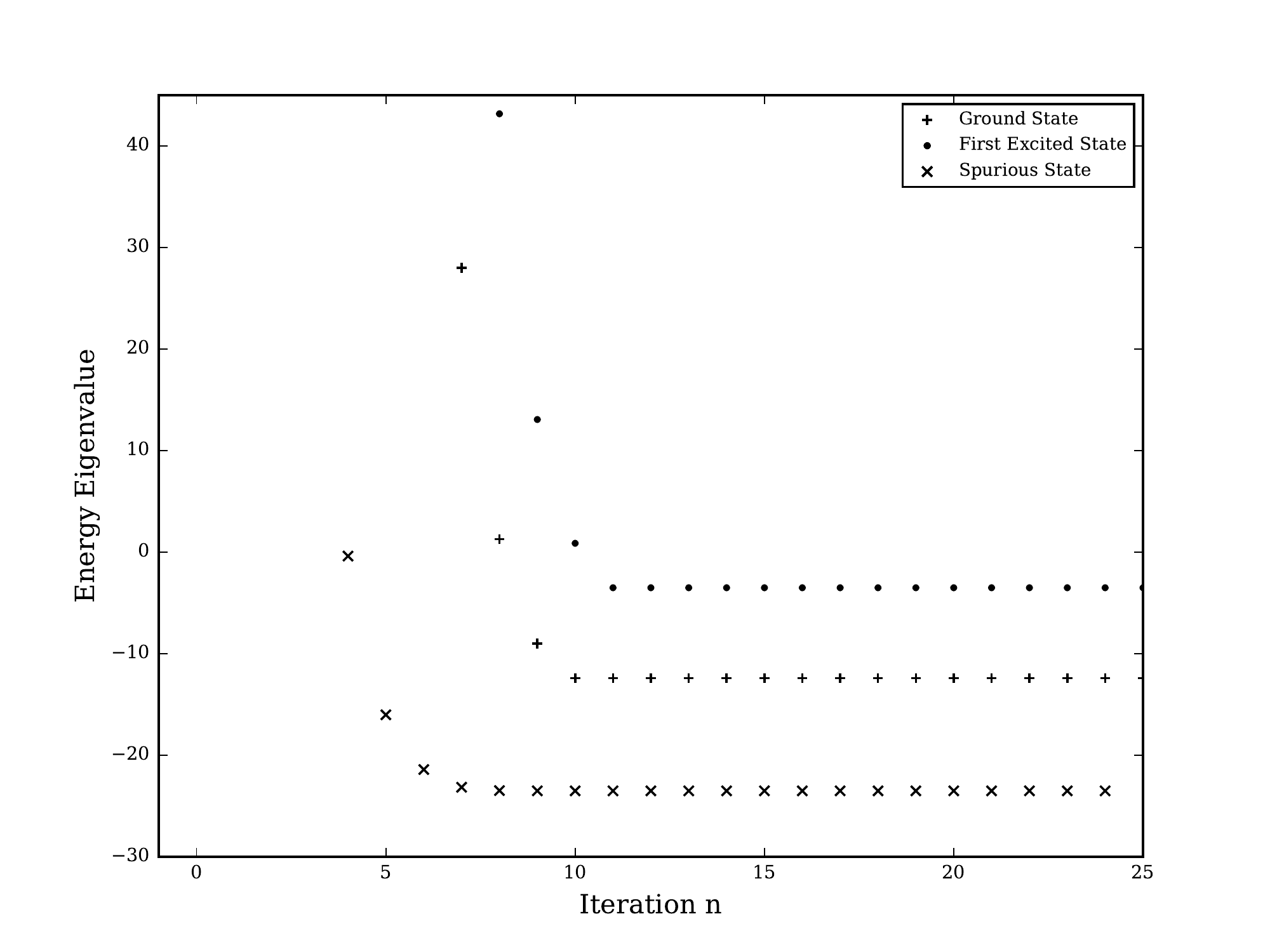}
     \caption{The iterative convergence behavior of the energy ground, excited, and spurious state eigenvalues obtained from the Lanczos algorithm.}
	\label{fig:eigenvalue_vs_iteration}
  \end{center}
\end{figure}
\begin{figure}[h!]
  \begin{center}
      \includegraphics[height=60mm]{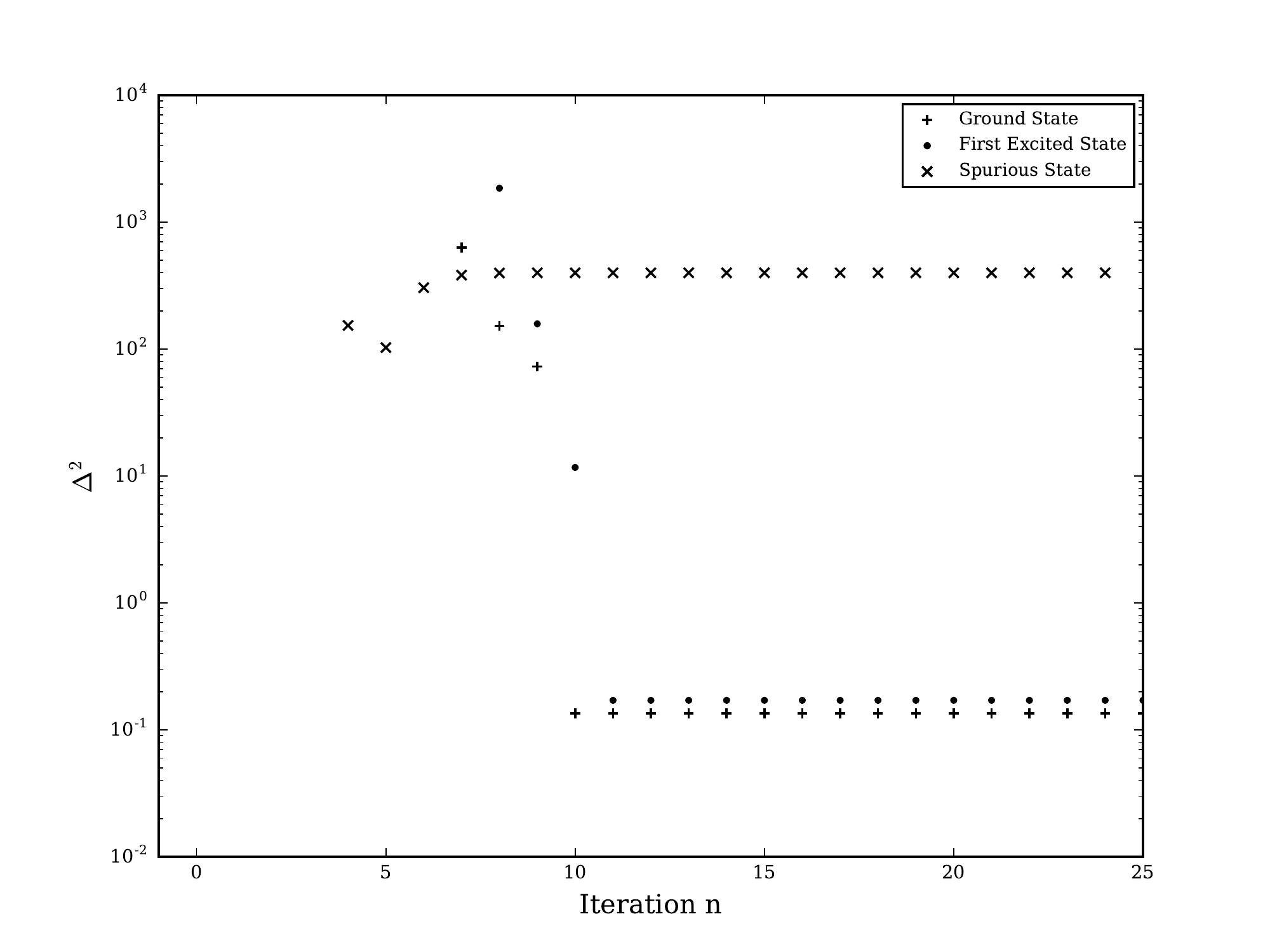}
     \caption{The iterative convergence behavior of $\Delta^{2}$ of the ground states, excited, and spurious state. Note that the ground and excited state converge to a relatively small value while the spurious state converges to a relatively large non-zero value.}
     \label{fig:delta_sq}
  \end{center}
\end{figure}

Choosing 30 basis states  the Lanczos algorithm is used to iteratively 
obtain approximate eigenstates of $\hat{H}$. Approximate energy  eigenvalues of the ground and excited states
were determined (see Fig \ref{fig:eigenvalue_vs_iteration}). 
After each iteration  $\Delta^{2}$ (see Fig. \ref{fig:delta_sq}) is calculated. It is clear that the ground state and excited state converge to relatively small values while the unphysical spurious state converges to a relatively large non-zero value. After $11$ iterations the approximate values of energy eigenvalues have converged to $e_{0,8}=12.397$ and $e_{2,11}=-3.579$ with a convergence factor of $\leq 10^{-3}$.

\begin{figure}[h!]
  \begin{center}
     \includegraphics[height=60mm]{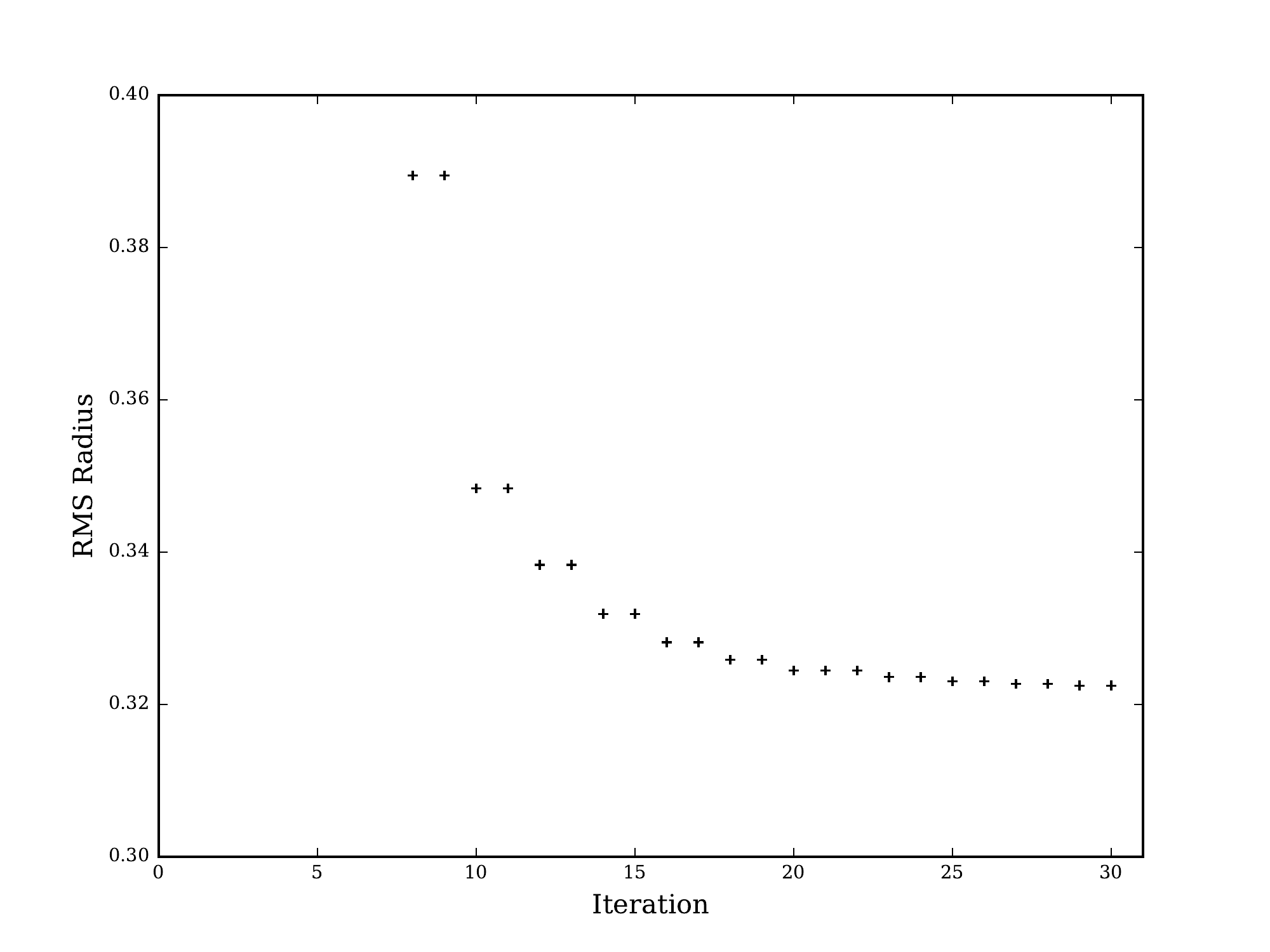}
     \caption{The convergence behavior of the RMS radius}
     \label{fig:RMS}
  \end{center}
\end{figure}

Having ascertained the ground state eigenfunction the rms radius
radius can easily be calculated (see Fig\ref{fig:RMS}). Note that the convergence is much slower than that of the eigenvalues. This is undoubtedly due to the fact that the convergence rate of the approximate ground state eigenfunction is considerably slower than that of the approximate ground state eigenvalue. After 30 iterations the RMS radius is $0.322$.

For a bare operator which contains a bound as well as a continuum spectrum the Lanczos algorithm has been used to obtain its approximate eigenstates. Unphysical spurious eigenstates which may appear can easily be identified. Since the approximate eigenfunctions are obtained the expectation value of other observables can easily be calculated without recourse to the use of effective operators.

\bibliography{lan2}
\end{document}